\documentclass[prd,floats,floatfix,showpacs,nofootinbib]{revtex4}
\usepackage{graphicx}
\usepackage{dcolumn}
\usepackage{bm}
\usepackage{graphics}
\usepackage{slashed}
\usepackage{amssymb}
\usepackage{natbib}
\usepackage{amsmath}
\usepackage{amsthm}
\usepackage{url}
\usepackage{color}
\newcommand{\fracc}[2]{\frac{\textstyle{#1}}{\textstyle{#2}}}

\begin{document}

\title{Chiral symmetry breaking as a geometrical process}

\author{E. Bittencourt$^{1,2,3}$}\email{eduardo.bittencourt@icranet.org}
\author{S. Faci$^{1}$}\email{sofiane@cbpf.br}
\author{M. Novello$^{1}$}\email{novello@cbpf.br}

\affiliation{$^{1}$Instituto de Cosmologia, Relatividade e Astrof\'isica ICRA - CBPF\\
Rua Doutor Xavier Sigaud, 150, Urca, CEP 22290-180, Rio de Janeiro, RJ, Brasil}
\affiliation{$^2$Sapienza Universit\`a di Roma - Dipartimento de Fisica\\
P.le Aldo Moro 5 - 00185 Rome - Italy}
\affiliation{$^3$ICRANet, Piazza della Repubblica 10 - 65122 Pescara - Italy}
\date{\today}

\begin{abstract}
This article expands for spinor fields the recently developed \textit{Dynamical Bridge} formalism which relates a linear dynamics in a curved space to a nonlinear dynamics in Minkowski space. Astonishingly, this leads to a new geometrical mechanism to generate a chiral symmetry breaking without mass, providing an alternative explanation for the undetected right-handed neutrinos. We consider a spinor field obeying the Dirac equation in an effective curved space constructed by its own currents. This way, both chiralities of the spinor field satisfy the same dynamics in the curved space. Subsequently, the dynamical equation is re-expressed in terms of the flat Minkowski space and then each chiral component behaves differently. The left-handed part of the spinor field satisfies the Dirac equation while the right-handed part is trapped by a Nambu-Jona-Lasinio (NJL) type potential.
\end{abstract}

\maketitle

\section{Introduction}
This work is related to the so-called \textit{Dynamical Bridge} formalism. The latter states that fundamental fields can satisfy equivalent dynamical equations in different background geometries. Namely, a given nonlinearity of the field in the flat space can be ``hidden" in the curvature of an effective curved space. This was shown to be true for spin $0$ and $1$ fields \cite{novetgoul,goulart,JCAP,nov_fal_gou,nov_bit} and can also be applied in the kinematical context \cite{nov_bit_gordon,nov_bit_drag}. Here we demonstrate that such procedure also holds for spinor fields. In particular, we show that the self-interacting term of the Nambu-Jona-Lasinio (NJL) equation--which appears commonly as a phenomenological regime in QCD based on chiral symmetry breaking (see historical details in Ref.\ \cite{hatsuda})--can be derived from the minimal coupling of a Dirac field with an effective curved geometry. Therefore, this Dynamical Bridge goes together with the chiral symmetry breaking even for massless spinor fields if the coupling constant is sufficiently large. We thus have a new and geometrical explanation for the non observation of the right-handed neutrinos.

In our proposal, we consider a massless spinor field $\Psi$ satisfying the Dirac equation $i \hat\gamma^\mu \hat\nabla_{\mu} \Psi = 0$ in a curved background without any gravitational character. This effective curved geometry represents the modification of the flat space caused only by the field $\Psi$. The latter modifies the space metric through
\begin{equation}
\label{first-eq}
\hat{g}^{\mu\nu}  =  \eta^{\mu\nu} + 2\,\alpha\, H^{\mu} H^{\nu},
\end{equation}
where the arbitrary function $\alpha$ depends on the scalars $A\equiv\bar\Psi \Psi$ and $B\equiv i \bar\Psi \gamma_5 \Psi$.
The vector $H^{\mu}$ is given in terms of the vector $J^{\mu}$ and axial $I^{\mu}$ currents. Afterwards we rewrite this dynamics in Minkowski space (equipped with the metric $\eta_{\mu\nu}$). This leads to the two following equations
\begin{eqnarray}
& i\gamma^\mu \partial_{\mu}\Psi_L=0,\\
& i\gamma^\mu \partial_{\mu}\Psi_R + s (A+iB\gamma_5) \, \Psi_L=0, \label{zzz}
\end{eqnarray}
where $s$ depends on $\alpha$ and its derivatives, $\Psi_L\equiv(1/2)({\bf 1}-\gamma_5)\Psi$ and $\Psi_R\equiv(1/2)({\bf 1}+\gamma_5)\Psi$ are respectively the left-handed and right-handed chirality components of $\Psi$, with $\Psi= \Psi_L+\Psi_R$. The left-handed component still obey the Dirac dynamics in Minkowski space while the right-handed component verifies the equation (\ref{zzz}), which is a generalization of the NJL dynamics. The curvature of the effective space goes to the nonlinearity of NJL dynamics in Minkowski space, but only for the right-handed component of the spinor field. This introduces a new answer to the experimentally observed chiral symmetry breaking of neutrinos and stands as an alternative for the Standard Model (SM) interpretation which assumes that right-handed neutrinos don't exist, i.e., they don't interact through the SM forces. Further extensions assume the right-handed neutrinos have very large masses which makes them invisible.

This article is organized as follows. The next section introduces the mathematical background to understand the Dynamical Bridge formalism and particularly how the curved space-time is built with spinors. Sec.\ [\ref{DB}] expands the Dynamical Bridge for spinor fields and exhibits how Dirac's dynamics in a curved space leads to a nonlinear dynamics in Minkowski space. In section [\ref{regime}] an important regime of the resulting nonlinear dynamics is explored yielding a generalized NJL equation.
Finally, Sec.\ [\ref{conclusion}] summarizes our concluding remarks.

\section{Mathematical Compendium for the Dynamical Bridge}
In order to properly construct the Dynamical Bridge for spinor fields let us describe some mathematical ingredients. It is well known that any metric tensor $\hat{g}^{\mu\nu}$ can always be decomposed into
\begin{equation}
\label{met}
\hat{g}^{\mu\nu} = g^{\mu\nu} + \Sigma^{\mu\nu},
\end{equation}
where $g^{\mu\nu}$ is some background metric and $\Sigma^{\mu\nu}$ is a rank two tensor field. Under this form the inverse metric $\hat{g}_{\mu\nu}$ should be expressed by an infinite series. However, the metric tensor $\hat g^{\mu\nu}$ admits an inverse with the same binomial form if we impose the condition $\Sigma^{\mu\nu} \, \Sigma_{\nu\lambda} = p \, \delta^{\mu}_{\lambda} + q \, \Sigma^{\mu}{}_{\lambda}$, where $p$ and $q$ are arbitrary functions of the coordinates. Throughout the text, we set the background metric $g_{\mu\nu}$ to be the Minkowski one $\eta_{\mu\nu}$ in arbitrary coordinate systems. Though this might appear restrictive, it can be generalized to arbitrary curved backgrounds \cite{spin_curv}.

With a generic spinor field $\Psi$ we construct two Hermitian scalars $A\equiv\bar\Psi \Psi$ and $B\equiv i \bar\Psi \gamma_5 \Psi$. We also define the associated vector and axial currents, respectively, as $J^{\mu} \equiv \bar\Psi_{} \gamma^{\mu} \Psi_{}$ and $I^{\mu} \equiv \bar\Psi_{} \gamma^{\mu} \gamma_5 \Psi_{}$. The $\gamma^{\mu}$'s are the Dirac matrices which satisfy the closure relation of the Clifford algebra
\begin{equation}
\label{clif_mink}
\{\gamma^{\mu},\gamma^{\nu}\} = 2 \, \eta^{\mu\nu} {\bf 1},
\end{equation}
where ${\bf 1}$ is the unity matrix of the algebra. This algebra must be valid for the effective metric $\hat g_{\mu\nu}$ and the background one as well. Using the Pauli-Kofink identity for an arbitrary element $Q$ of the Clifford algebra
\begin{equation}\label{pkofink}
(\bar\Psi Q\gamma_{\lambda}\Psi)\gamma^{\lambda}\Psi=(\bar\Psi Q\Psi)\Psi - (\bar\Psi Q\gamma_5\Psi)\gamma_5\Psi,
\end{equation}
the following relations are easily derived
\begin{equation}
\label{rel_sc_cur}
J^2 = - I^2 = A^2+B^2, \qquad J_{\mu}I^{\mu} = 0,
\end{equation}
where we have denoted $X^2\equiv \eta_{\mu\nu} \,  X^{\mu}X^{\nu}$ for the vectorial objects.

The Dynamical Bridge method consists in writing the term $\Sigma^{\mu\nu}$ of the curved space-time metric (\ref{met}) in terms of the dynamical field, which is given by the spinor $\Psi$ in the present work. The simplest way to do this is to set
\begin{equation}
\label{Sigma2}
\Sigma^{\mu\nu} \doteq 2\,\alpha\, H^{\mu} H^{\nu},
\end{equation}
where $H^{\mu} \doteq J^{\mu} + \epsilon I^{\mu}$ is a linear combination of the currents, $\epsilon$ is an arbitrary constant and $\alpha$ is an arbitrary function of the scalars $A$ and $B$. The factor $2$ is conventional. Eqs.\ (\ref{pkofink}) and (\ref{rel_sc_cur}) leads to the following identities
\begin{equation}
\label{gamma.H}
H_{\mu}\gamma^{\mu}\,\Psi = ({\bf 1}+\epsilon\gamma_5) (A+iB\gamma_{5})\Psi, \qquad \mbox{and} \qquad H^2=(1-\epsilon^2)J^2.
\end{equation}
Finally, using the relation $\hat g^{\mu\nu} \hat g_{\nu\lambda} = \delta^\mu_{\ \lambda}$, the metric tensor and its inverse can be written as
\begin{equation}\label{hatg}
\begin{array}{l}
\hat g^{\mu\nu} = \eta^{\mu\nu} + 2\, \alpha \, H^{\mu} H^{\nu},\\[2ex]
\hat g_{\mu\nu} = \eta_{\mu\nu} -  \fracc{2\, \alpha}{1+ 2\, \alpha H^2} \, H_{\mu} H_{\nu}.
\end{array}
\end{equation}

From now on, we make use of the tetrad formalism \cite{tetrad} (or vierbeins), in order to perform the Dynamical Bridge by changing the tangent (spinor) space and the physical space-time simultaneously. This allows to simplify the calculations since we are dealing with spinors in curved background. The Lagrangian and the dynamical equations for $\Psi$ must then be invariant not only under diffeomorphisms, but also under local Lorentz transformations acting on the tangent space.

Let us now rewrite the objects defined previously in terms of the tetrads. We define two tetrad bases $e^{\mu}{}_A$ and $\hat{e}^{\mu}{}_A$ which relate the tangent space provided with the metric $\eta_{AB}$ to the physical spaces endowed with the two metrics $\eta_{\mu\nu}$ and $\hat g_{\mu\nu}$. The two bases satisfy the following relations
\begin{equation}
\label{tetrad}
\begin{array}{l}
\hat{g}^{\mu\nu} = \eta^{AB} \, \hat e^{\mu}{}_{A} \, \hat e^{\nu}{}_{B},\\[2ex]
\eta^{\mu\nu} = \eta^{AB} \, e^{\mu}{}_{A} \, e^{\nu}{}_{B}.
\end{array}
\end{equation}
The Greek indices refer to the physical spaces and are lowered and raised by the corresponding metric ($\eta_{\mu\nu}$ or $\hat g_{\mu\nu}$). The capital Latin indices refer to the tangent space and are lowered and raised by $\eta_{AB}=\mbox{diag}(1,-1,-1,-1)$. The inverse tetrad bases $e_{\mu}{}^{A}$ and $\hat{e}_{\mu}{}^{A}$ should satisfy $ e_{\mu}{}^{A} \, e^{\nu}{}_{A} = \hat e_{\mu}{}^{A} \,\hat e^{\nu}{}_{A} = \delta_{\mu}^\nu$ and $e_{\mu}{}^{A} \, e^{\mu}{}_{B}= \hat e_{\mu}{}^{A} \,\hat e^{\mu}{}_{B}= \delta^{A}_{B}$. Furthermore, any vector $X^\mu$ (or $\hat X^\mu$) in the space-time $\eta_{\mu\nu}$ (or $\hat g_{\mu\nu}$) has a counterpart $X^A\equiv e^A_\mu  X^\mu$ (or $\hat X^A\equiv\hat e^A_\mu \hat X^\mu$) in the tangent space. In particular, for the Dirac matrices we assume that
\begin{equation}
\gamma^A = \hat{e}_{\mu}{}^{A} \, \hat\gamma^\mu= e_{\mu}{}^{A} \, \gamma^\mu,
\end{equation}
where $\gamma^A$'s are the constant Dirac matrices.
Note that the Dirac matrices $\hat\gamma^\mu$ also verify the closure relation of the Clifford algebra but the one related to the effective curved space
\begin{equation}
\{\hat \gamma^{\mu}, \hat \gamma^{\nu}\} = 2 \, \hat g^{\mu\nu}\, {\bf 1}.
\end{equation}

For consistency with Eqs.\ (\ref{hatg}), the two tetrad bases must obey the condition
\begin{equation}
\label{mapa_tetr}
\hat e^{\mu}{}_{A}= e^{\mu}{}_{A} + \beta \, H_{A} H^\mu
\end{equation}
and, using Eqs.\ (\ref{hatg}) and (\ref{tetrad}), we have the constraint $\alpha =  2\beta/(2+\beta H^2)$. Therefore, the inverse tetrad bases are related according with
\begin{equation}
\label{inv_mapa_tetr}
\hat e_{\mu}{}^{A}= e_{\mu}{}^{A} - \fracc{\beta}{1+\beta H^2} \, H^{A} H_\mu.
\end{equation}
That is all we need about spinor fields in curved spaces and tetrad formalism in order to construct the bridge between the dynamics. Next section we shall focus on the map itself.

\section{Dynamical Bridge for spinor fields: From Dirac to a nonlinear dynamics}\label{DB}
Now we describe how the Dynamical Bridge works for spinor fields. As mentioned before, we start with the linear Dirac equation in an effective curved background, then we expand the formulas to reach a new and different, though mathematically equivalent, dynamics in the Minkowski background. The resulting equation is a generalization of the NJL dynamics.

In the effective curved geometry given by (\ref{hatg}), the Dirac equation for the spinor field\footnote{Unlike what was done in Ref.\ \cite{novdirac}, here $\Psi$ remains the same in both spaces.} reads
\begin{equation}
\label{Dirac}
i\hat\gamma^\mu\hat\nabla_{\mu}\Psi=0,
\end{equation}
where $\hat\nabla\equiv\partial_{\mu} - \hat\Gamma_{\mu}$ and the Fock-Ivanenko connection $\hat\Gamma_{\mu}$ is given by
\begin{equation}
\label{christoffel}
\hat\Gamma_{\mu} = -\frac{1}{8}\left( [\hat\gamma^{\alpha},\hat\gamma_{\alpha,\mu}] - \hat \Gamma^{\rho}_{\alpha\mu}[\hat\gamma^{\alpha},\hat\gamma_{\rho}] \right).
\end{equation}
Note that a comma means partial derivative, $\hat \Gamma^{\rho}_{\alpha\mu}$ is the Christoffel symbol constructed with $\hat g_{\mu\nu}$ and the squared brackets represent the commutator operator. Introducing the tetrads allows to rewrite the Fock-Ivanenko connection (\ref{christoffel}) as
\begin{equation}
\hat\Gamma_{\mu} = \hat e_\mu{}^A \hat\Gamma_{A}, \qquad \mbox{with}\qquad \hat\Gamma_{A}=-\frac{1}{8}\,\hat\gamma_{BCA}[\gamma^{B},\gamma^{C}],
\end{equation}
where $\hat\gamma_{ABC}$ is called spin connection and
\begin{equation}
\label{C_ABC}
\hat\gamma_{ABC} = \frac{1}{2} (\hat C_{ABC} - \hat C_{BCA} - \hat C_{CAB}) \quad \mbox{and} \quad
\hat C_{ABC} = - \hat e_{\nu A} \hat e^{\mu}{}_{[B} \hat e^{\nu}{}_{C], \mu}.
\end{equation}
It is easy to see that $\hat C_{ABC} = - \hat C_{ACB}$ and, consequently, $\hat\gamma _{ABC} = - \hat\gamma_{BAC}$. To go further we use the relation (\ref{mapa_tetr}) to expand (\ref{C_ABC}), as follows
\begin{equation}
\label{C_ABC_Mink}
\hat C_{ABC}= -\fracc{1}{1+\beta H^2}(H_{A}\beta_{,[B} H_{C]}+ \beta^2 H_A H_{[B}\dot H_{C]} - \beta \, H_{A}H_{[C,B]}) +\beta H_{A,[B}H_{C]}-\fracc{\beta^2}{2(1+\beta H^2)}H_{A}H^2_{,[B}H_{C]},
\end{equation}
where $\dot H_A$ holds for $H_{A,B}H^B$. Thus, we compute the spin connection
\begin{equation}
\label{gamma_ACB_Mink}
\hat \gamma_{ABC} = \hat C_{ABC} +\frac{\beta}{2}\left[\fracc{\beta H^2}{1+\beta H^2}(H_A H_{[B,C]}+ H_{B}H_{[C,A]}-H_{C}H_{[B,A]}) + H_{A,[C}H_{B]} + H_{[B,C]} H_A\right].
\end{equation}

The Dirac equation (\ref{Dirac}) reads
\begin{equation}
\label{tetr_Dirac}
i\hat\gamma^A(\hat\partial_{A} - \hat\Gamma_{A})\Psi \ = \ i\gamma^A\left(\partial_A+\beta H_AH^B\partial_B +\frac{1}{8}\hat\gamma_{BCA}[\gamma^{B},\gamma^{C}]\right)\Psi=0,
\end{equation}
where $\hat \partial_A\equiv\hat e^\mu{}_{A} \partial_{\mu}= \partial_A +\beta H_AH^B\partial_B$.
To proceed we need to compute the expression $\hat\gamma_{BCA}\gamma^A\gamma^B\gamma^C$. This turns out to contain only two non-vanishing terms\footnote{To do this, we use the following algebraic identity for the Dirac matrices $\gamma^A\gamma^B\gamma^C = \eta^{AB}\gamma^C + \eta^{BC}\gamma^A - \eta^{AC}\gamma^B + i\epsilon^{ABC}{}_{D}\gamma^D\gamma_5$.}, which are
\begin{equation}
\label{int1}
\hat\gamma_{BCA}\epsilon^{ABC}{}_{D}=\fracc{\beta^2 H^2}{2(1+\beta H^2)} H_A H_{[B,C]} \epsilon^{ABC}{}_{D}
\end{equation}
and
\begin{equation}
\label{int2}
\begin{array}{lcl}
\hat\gamma_{BCA}\eta^{AB}\gamma^C&=&\fracc{1}{1+\beta H^2}\left[\dot\beta H_A\gamma^A- H^2(\beta .\gamma) +\beta^2 H^2 \dot H_A \gamma^A-\beta^2 \dot H_A\,H^A H_B\gamma^B -\fracc{\beta}{4}(H^2 . \gamma)+\fracc{\beta}{2}\dot H_A \gamma^A\right]\\[2ex] &&-\fracc{\beta}{4}(H^2 . \gamma) - \fracc{\beta^2}{2(1+\beta H^2)} \dot{H^2} H_A\gamma^A + \fracc{\beta^2}{2(1+\beta H^2)}H^2(H^2.\gamma).
\end{array}
\end{equation}
We have used the shortcuts $\dot\beta\equiv\beta_{,A}H^A$, $\beta.\gamma\equiv \beta_{,A}\gamma^A$ and $H^2.\gamma\equiv2H^AH_{A,B}\gamma^B$. Eq.\ (\ref{tetr_Dirac}) does not match any well-known dynamic for a nonlinear spinor field in flat space. To get a simpler equation, we should assume a specific value for $\epsilon$. From Eq.\ (\ref{gamma.H}), setting $\epsilon^2=1$, it implies that $H^2=0$ drastically simplify the dynamical equations\footnote{It should be remarked that this does not imply $J^2=0$ which otherwise would be too restrictive.}. Let us assume $\epsilon=-1$ for later convenience. Consequently, the expression (\ref{int1}) vanishes and (\ref{int2}) reduces to
\begin{equation}
\label{int2_simpl}
\hat\gamma_{BCA}\eta^{AB}\gamma^C=\dot\beta H .\gamma+\frac{\beta}{2}\dot H.\gamma.
\end{equation}
Therefore, the spinor field equation (\ref{Dirac})  written in the curved space without any gravitational character, becomes the following equation in the flat space
\begin{equation}
\label{dirac_simpl}
i\left(\gamma^{\mu} \partial_{\mu} +\beta\gamma^{\mu} H_{\mu} H^{\nu} \partial_{\nu} +\frac{\beta}{4}\dot H_{\mu}\gamma^{\mu} + \frac{\dot\beta}{2} H_{\mu}\gamma^{\mu}\right) \, \Psi =0.
\end{equation}
This is a nonlinear dynamical equation for the spinor field in the Minkowski space. The terms originated from the effective curved space connection are interpreted as self-interacting terms. As it is, this equation does not fit any known nonlinear dynamics for spinor fields \cite{fushchych}. Nonetheless more simplifications can be applied and some remarkable results emerge when we examine closely the coupling parameter $\beta$ and look for particular solutions.

\section{Breaking chiral symmetry without mass}\label{regime}
In the standard model of particles, the neutrino is represented by a massless spinor field. It means that the results of the Dynamical Bridge developed above are particularly applicable for the scenarios where this particle is involved. In this vein, let us remark that in the Lagrangian of the standard model, the right-handed neutrinos are not present since weak interactions couple only to the left-handed neutrinos and due to the symmetries of the model a mass term for neutrinos is not allowed. Yet the 1998's experiment, Super-Kamiokande, and various other independent and different experiments have detected the family oscillation of this particle \cite{pdg}. However, from the minimal coupling principle applied to Minkowski space, this can be explained if the neutrinos are lightly massive, breaking the chiral symmetry. But a fundamental question raises: where are the not yet observed right-handed neutrinos?

Nowadays, there are two possible ways to answer this question: either they just do not exist and then massive neutrinos are of Majorana type, which is not really satisfactory; or, they exist but do not interact weakly. We should also remark that due to the mixing of chiralities caused by the mass term, it was led to assume that right-handed neutrinos are heavy enough in order to hide them from observations. Then, the chiral symmetry is broken ``by hand" and it does not really explain why such neutrinos are so massive. Several experiments have unsuccessfully tried direct or indirect detection of such neutrinos, as one can see in \cite{drewes,Canetti} and references therein.

Notwithstanding, this work provides a third possible explanation for the chiral symmetry breaking according to a geometrical viewpoint. Indeed, in the regime $\dot\beta\gg\beta$ the Eq.\ (\ref{dirac_simpl}) reduces to
\begin{equation}
\label{dirac_simpl_alg}
i\left(\gamma^{\mu} \partial_{\mu} + \frac{\dot\beta}{2} H_{\mu}\gamma^{\mu}\right) \, \Psi =0.
\end{equation}
Substituting the expression for $H_{\mu}\gamma^{\mu}\,\Psi$ from Eq.\ (\ref{gamma.H}), we obtain
\begin{equation}
\label{dirac_simpl_ext}
i\gamma^{\mu} \partial_{\mu}\Psi + i\frac{\dot \beta}{2}({\bf 1}-\gamma_5)(A+iB\gamma_5) \, \Psi =0.
\end{equation}
This equation is very similar to the NJL equation for a spinor field propagating in Minkowski space-time. This means that the nonlinear self-interacting term can be interpreted as a modification of the space-time structure. Up to this point, our method seems to present two equivalent dynamical equations though written in two different spaces. But there is a ``hidden" property that appears in the flat case when we decompose the spinor $\Psi$ into its two opposite chiralities. That is
$$\Psi=\Psi_L+\Psi_R=\frac{1}{2}({\bf 1}-\gamma_5)\Psi+\frac{1}{2}({\bf 1}+\gamma_5)\Psi,$$
where $\Psi_L$ and $\Psi_R$ are components of the spinor field representing the left and the right-handed chiralities. Therefore, Eq (\ref{dirac_simpl_ext}) splits into two distinct parts
\begin{eqnarray}
& i\gamma^\mu \partial_{\mu}\Psi_L=0,\\
& i\gamma^\mu \partial_{\mu}\Psi_R + i\dot\beta (A+iB\gamma_5) \, \Psi_L=0.
\end{eqnarray}

From these equations, it follows the remarkable result: each chiral component $\Psi_L$ and $\Psi_R$ satisfies a different dynamical equation in the Minkowski space. The left-handed component propagates as a free Dirac field when the right-handed component is trapped by the self-interacting term. If the coupling parameter $\dot \beta$ is sufficiently large, the right-handed component needs very high energies to be detected. This leads to a new and geometrical explanation for the non observation of the right-handed neutrinos.

\section{Concluding remarks}\label{conclusion}
This paper completes a sequence of works dealing with the \textit{Dynamical Bridge}, a mathematical equivalence between different dynamical equations of a given field (scalar, vector and now spinor) defined in different space-times. In particular, we have shown that a massless spinor field satisfying the linear Dirac equation defined in a curved geometry is equivalent to the same spinor field in the Minkowski space with its right-handed component satisfying a modified NJL equation and the left-handed component still satisfying the massless Dirac equation. If the coupling parameter is sufficiently large, the right-handed component needs very high energies to be detected. This leads to a new and geometrical explanation for the non observation of the right-handed neutrinos.

It is important to notice that the set of solutions of Eq.\ (\ref{dirac_simpl}) is not empty. We have at least one special and interesting class of $\Psi$'s provided by the Inomata \cite{Inomata} condition $\Psi_{, \mu} = -(1/2)\dot \beta H_\mu \Psi$ which satisfies this equation. Moreover, this class satisfies also Eq.\ (\ref{dirac_simpl_alg}) and then leads to chiral symmetry breaking for all regimes of $\beta$, resulting in an explicit solution.

In the case of massive fermions, we expect that each fermion possesses a characteristic energy scale in which this nonlinear coupling is relevant, i.e., $\beta=\beta(m)$ where $m$ is the mass. Thus, this kind of self-interaction does not appear at low energies and the chiral symmetry should be broken by the mass term. On the other hand, the standard model takes into account self-interaction only as quantum corrections. Then, it should be carefully examined if quantum corrections could model such generalized NJL self-interaction and if it would be distinguished from our proposal. Nowadays there are only phenomenological properties about the right-handed neutrinos \cite{drewes} and no observation has been successful. The attempt to introduce specific values to this model is postponed to a more experimental work.

Note that, as a drawback, there is no conservation of the probability current of the spinor field obtained in the Minkowski space through the Dynamical Bridge. This can be explained by two different ways: the spinor field perturbations propagate along geodesics in the curved space but not in Minkowski space; this case is then equivalent to the scalar one \cite{goulart}. The second possible explanation states that the chiral symmetry breaking does also break the unitarity of the Dirac theory. To avoid this difficulty, another proposal is possible by changing the form of the curved geometry. Instead of Eq.\ (\ref{first-eq}), if we consider an effective metric of the form
\begin{equation}
\label{mod_met}
\hat{g}^{\mu\nu} =  \alpha_1 g^{\mu\nu} + 2\,\alpha_2 \, H^{\mu} H^{\nu},
\end{equation}
where $\alpha_1\neq 1$ and $\alpha_2$ are initially two arbitrary and independent functions of the scalars $A$ and $B$. The current is now conserved in both the effective curved and flat spaces provided $\alpha_1=\alpha_1(\alpha_2)$; that is only one parameter is free. However, preliminary attempts indicate that the dynamics in this case is no longer given by the NJL models. So, this would deviate from our original proposal and should be discussed in a future work.

\section*{Acknowledgments}
We acknowledge the participants of the ``Pequeno Semin\'ario" for their comments. This work was supported by CNPq, CAPES (BEX 13956/13-2) and FAPERJ of Brazil.

\end{document}